\documentclass[12pt]{iopart}

\usepackage{pdfpages} 
\usepackage{graphicx} 
\usepackage{amssymb} 
\usepackage{amsbsy}
\usepackage{url}
\def\R{\mathbb{R}}

\newcommand{\codes}[1]{\textsf{#1}}
\newcommand{\pounders}{\codes{POUNDERS}}

\newcommand{\tao}{\codes{TAO}}
\newcommand{\xh}{\hat{\mathbf{x}}}
\newcommand{\same}[1]{\textbf{#1}}
\newcommand{\xb}{\mathbf{x}}
\newcommand{\db}{\mathbf{d}}
\newcommand{\lb}{\mathbf{l}}
\newcommand{\ub}{\mathbf{u}}
\newcommand{\svb}{\mathbf{s}}
\newcommand{\Fb}{\mathbf{F}}
\newcommand{\nub}{\boldsymbol \nu}
\newcommand{\sigb}{\boldsymbol \sigma}
\newcommand{\eb}{\boldsymbol \epsilon}

\begin{document}
\title[DFO for NP]{Derivative-free optimization for parameter
estimation in computational nuclear physics}

\author{Stefan M Wild,$^1$ Jason Sarich,$^1$ and Nicolas Schunck$^2$}

\address{$^1$ Mathematics and Computer Science Division,
Argonne National Laboratory, Argonne, IL 60439, USA}
\address{$^2$ Physics Division, Lawrence Livermore National Laboratory,
Livermore, CA, 94551, USA}

\ead{wild@anl.gov}

\begin{abstract}
We consider optimization problems that arise when estimating a
set of unknown parameters from experimental data,
particularly in the context of nuclear density functional theory. We examine the cost
of not having derivatives of these functionals with respect to the parameters.
We show that the \pounders\ code for local derivative-free optimization obtains
consistent solutions
on a variety of computationally expensive energy density functional calibration
problems.
We also provide a primer on the operation of the \pounders\ software
in the Toolkit for Advanced Optimization.
\end{abstract}

\submitto{\JPG}
\maketitle

\section{Introduction}
\label{sec:intro}

Much of the intellectual capital in nuclear physics is invested in
\emph{forward problems} whereby a theory or model is posited, assumptions are
added (to improve accuracy) and relaxed (to improve universality), and
hypotheses are tested. An example can be seen with the \emph{ab
initio} approach to nuclear structure.  Here, the form of the Hamiltonian is
derived from chiral effective field theory \cite{epelbaum2012}.
One of the basic assumptions is that the nuclear
many-body problem can be solved non-relativistically with nucleons as basic
degrees of freedom \cite{barrett2013}.
The truncation in chiral perturbation and
the inclusion or neglect of three- and $N$-body forces are some of the
hypotheses that can be tested by comparing model predictions with
experimental data \cite{navratil2007}. Today, work on forward problems
invariably extends along a computational axis as well: models are made
computationally tractable and numerically implemented, and computational
performance and efficiency are improved.

Equally important is the \emph{inverse problem}: given data (experimental or
otherwise) and a forward model, free parameters for the model are determined
based on the data. This aspect is especially important in the context of the
nuclear shell model or nuclear density functional theory (DFT). Indeed, these
approaches to the nuclear many-body problem are a notch more phenomenological
than \emph{ab initio} theory: they rely on an effective interaction, or
alternatively an effective energy density, that is not predetermined from some
underlying theory \cite{bender2003,ring2000}. Obtaining a robust and reliable
estimate of the free parameters is essential since nuclear DFT is widely used
in a number of applications, from large-scale surveys of nuclear properties
\cite{erler2012} to fission \cite{younes2011}, and will play a critical role in
the physics explored at the future Facility for Radioactive Ion Beams
\cite{balantekin2014}. In this paper we focus on numerical optimization, one
aspect of inverse problems that often presents a bottleneck when working with
computationally expensive forward models.

Formally, we assume a collection of $n_d$
components of scalar data $\db=(d_1,\ldots, d_{n_d})$ based on which we must
determine values of $n_x$ real parameters $\xb=(x_1,\ldots, x_{n_x})$. It is
often convenient to think of a model $m$ as generating the observable $d_i$ based
on the set of real parameters $\xb$ and a set of hyperparameter values,
$\nub_i\in \R^p$, which represent known values needed to compute the forward
problem (such as the number of protons and the number of neutrons).
Thus the inverse problem is to determine the value(s) $\xb_*$ such that
\begin{equation}
m\left(\xb_*; \nub_i\right) \approx d_i \qquad i=1, \ldots, n_d.
 \label{eq:inverse}
\end{equation}
The level of agreement dictated by ``$\approx$'' can depend on the
uncertainties in the model $m$, the parameters $\xb_*$, and/or the data $\db$.

Parameter estimation typically depends on the distribution of the errors
between reality and the data. Given an assumed distribution of these errors, a
common approach in both Bayesian and frequentist parameter estimation is to
determine the maximum likelihood estimate (or maximum \textit{a posteriori}
estimate for Bayesians) for the parameters, namely, those
values that, given the values of the data, are most likely under the assumed
distribution(s).

Regardless of the distribution or whether the errors are independent of one
another, one generally arrives at an optimization problem. For example, if the
model $m$ is correct, the errors are independent, and the errors are Gaussian
with mean zero and known variance $w^2_i>0$, then maximizing the
log-likelihood (and hence the likelihood) is equivalent to solving
\begin{equation}
\min_{\xb\in \R^{n_x}} \sum_{i=1}^{n_d} \left(\frac{m\left(\xb; \nub_i \right)
- d_i}{w_i}\right)^2.
 \label{eq:gerrors}
\end{equation}
If the errors are correlated, then (\ref{eq:gerrors}) becomes
\begin{equation}
\min_{\xb\in \R^{n_x}} \sum_{i=1}^{n_d}\sum_{j=1}^{n_d} w_{ij}
\left(m\left(\xb; \nub_i\right) - d_i\right)\left(m\left(\xb; \nub_j\right)
- d_j\right),
 \label{eq:gderrors}
\end{equation}
where $w_{ij}$ captures the (inverse) covariance between the errors of
observables $i$ and $j$.

The objective in (\ref{eq:gerrors}) differs from $\chi^2$ objectives by a
constant factor (related to the degrees of freedom, $n_d-n_x$), and hence the
solution of (\ref{eq:gerrors}) with an appropriate $w$ (see \cite{DNR14})
arises throughout computational science.
Similar objective functions to be optimized can be
derived for a wide variety of other distributions,  including cases
where the variances $\{w_i^2:i=1, \ldots, n_d\}$ are unknown or specified only
by
a diffuse prior. These latter cases are especially relevant to nuclear DFT,
since there is little {\it a priori} information about the errors on computed observables.
Likewise, if constraints on the parameters are imposed (e.g.,
to break symmetries or satisfy physical realities), the optimization problem
can be modified to consider the restriction $\xb\in \Omega \subset \R^{n_x}$.

As we will see, the derivatives $\frac{\partial }{\partial x_j}m\left(\xb;
\nub_i\right)$ play a crucial role in identifying solutions to such
optimization problems. The solution of these problems is especially
difficult when such derivatives are not made available to the optimization
solver; such ``derivative-free'' situations are pervasive when evaluating
$m(\cdot; \cdot)$ entails running a legacy computer simulation.
In Section~\ref{sec:methods} we review methods for solving problems of the form
(\ref{eq:gerrors}) in both the unconstrained and bound-constrained case. We
 focus on derivative-free approaches for calibrating energy
density functionals and review the Practical Optimization Using No Derivatives
for sums of Squares (\pounders) method for solving such problems.
In Section~\ref{sec:starting}, we examine some of the optimization problems
from the UNEDF0, UNEDF1, and UNEDF2 parameterizations \cite{UNEDF0,UNEDF1,UNEDF2}. We show that
despite the potentially multimodal nature of the objective function
considered, the solutions obtained by \pounders\ are surprisingly robust to the
choice of starting point.
Our results also offer a further empirical validation of the sensitivity
analysis conducted in the UNEDF studies.
Section~\ref{sec:ders} returns to the matter of derivatives. Through a specific
example involving nuclear masses, we show that the availability of
derivatives with respect to even a few parameters can improve the efficiency of
the optimization.
In Section~\ref{sec:usage} we provide details on the usage of the \pounders\ 
method as well as general tips for solving such problems.
Section~\ref{sec:summary} concludes the paper.

\section{Optimization-Based Approaches for Parameter Estimation}
\label{sec:methods}

We will restrict our focus to (\ref{eq:gerrors}), the most common form  of
optimization problem encountered in parameter estimation, but we note that much of
our discussion applies for more general objectives. In many practical
applications, in particular the optimization of energy densities in DFT, the
model $m$ is a nonlinear function of the parameters $\xb$; hence the problem in
(\ref{eq:gerrors}) is that of \emph{nonlinear least squares} (NLS),
\begin{equation}
 \min_{\xb\in \Omega} f(x) = \sum_{i=1}^{n_d} F_i ( \xb )^2,
 \label{eq:nls}
\end{equation}
where the vector mapping $\Fb:\R^{n_x} \to \R^{n_d}$ captures the weighted
residuals and $\Omega$ can correspond either to all of $\R^{n_d}$ (called the
``unconstrained'' case) or to some subset of $\R^{n_d}$ (e.g., when
non-negativity, $x_i \geq 0$, is imposed for some parameter
$x_i$).

Solutions to (\ref{eq:nls}) are referred to as \emph{global} minimizers, and
such points $\xh$ have the property that $f(\xh)\leq f(\xb)$ for all $\xb\in
\Omega$. However, finding global solutions for arbitrary functions $\Fb$ is
generally intractable. Consequently, optimization methods that promise global
solutions are either making problem-specific assumptions (e.g., that $\Fb$ is a
linear function of $\xb$ or that $\Omega$ contains a finite number of points),
guaranteeing global optimality only asymptotically (and thus never achieved in
practice), or overstating their claims.

\begin{figure}[tb]
\begin{center}
\includegraphics[width=0.45\linewidth]{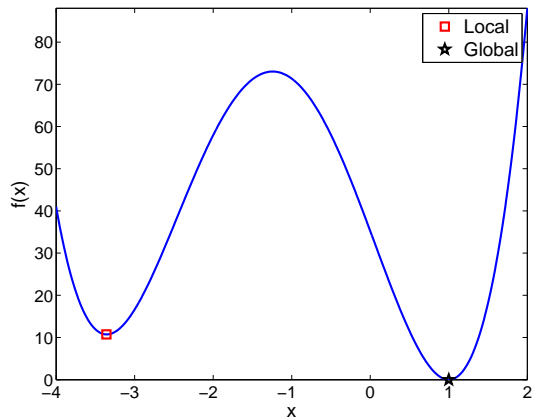}
\caption{Example of the one-dimensional nonlinear least squares problem
$f(x) = \sum_{i=1}^{n_{d}} F_{i}(x)^{2} $ with $n_{d} = 3$,
$F_i(x)=(x-\nu_i)^2-d_i$,
$\nub=(1, \, 1.1, \, 1.5)$, and $\db=(1-\nub)^2$. }
\label{fig:multimode}
\end{center}
\end{figure}

As a result, we follow the approach of seeking \emph{local} minimizers, which
cannot be improved upon locally: $f(\xh)\leq f(\xb)$ for all $\xb\in
\Omega$ close to $\xh$.
Figure~\ref{fig:multimode} illustrates that even simple, one-dimensional ($n_x=1$)
NLS problems can have multiple local minimizers, with
potentially all but one of these being nonglobal minimizers.
Hence, one must apply local optimization methods to such problems with caution;
Section~\ref{sec:starting} returns to this topic.

\subsection{Derivatives and Methods for Nonlinear Least Squares}
When the residual vector $\Fb$ is differentiable, the gradient of $f$ with
respect to the parameters $\xb$ is $\nabla_\xb
f(\xb) =2\sum_{i=1}^{n_d} F_i(\xb) \nabla_{\xb} F_i ( \xb )$ and plays a
crucial role in local optimality conditions.
In the unconstrained case, a necessary condition for
local minimizers is that the gradient of the function disappear, $\nabla_\xb
f(\xb) = 0$. In the constrained case, things are slightly more complex. Here,
we focused on one of the simplest cases, when bound constraints
\begin{equation}
 \Omega=\{\xb\in \R^{n_x}: l_i \leq x_i \leq u_i , i =1,\ldots, n_x\}
 \label{eq:bounds}
\end{equation}
are the only ones present. In the bound-constrained case, a
necessary condition is that $\xh\in \Omega$ and that the components of the
gradient satisfy
\[\frac{\partial f(\xh)}{\partial x_i} \quad  \left\{ \begin{array}{ll}
                                 =0 & \mbox{if } l_i < \hat{x}_i < u_i \\
                                 \geq 0 & \mbox{if } \hat{x}_i = l_i \\
                                 \leq 0 & \mbox{if } \hat{x}_i = u_i \\
                                 \end{array}
                                 \right.  \qquad i=1, \ldots , n_x.
\]
We say that a bound (or parameter, in this case) is ``active'' if the parameter
attains the bound (e.g., $\hat{x}_i=l_i$ or  $\hat{x}_i=u_i$).

In both the unconstrained and constrained cases, the derivatives
$\nabla_\xb f$ (and hence $\nabla_{\xb} F_i$)
play a vital role in guaranteeing decrease of the objective $f$, accelerating
convergence, and recognizing a solution. In most practical problems, the
residual $F_i(\xb)$ invariably depends on the output of a numerical or physical
simulation, and hence such derivatives may not readily be available.

When
these residuals are defined by a computer code free of proprietary
libraries and control flow logic that may introduce discontinuities,
\emph{algorithmic differentiation (AD)} \cite{Bischof2008} can be an invaluable
technique. AD tools generate source code---often automatically---by propagating
the chain rule through the original code. Under infinite-precision arithmetic,
derivatives from AD are exact.
Alternatively, one can apply
numerical differentiation (ND) to obtain approximate derivatives.
With ND, however, one
must take great care in selecting an appropriate finite-difference
stepsize on noisy simulations \cite{more2011edn}; also, the cost of
obtaining a full gradient using ND is generally at least $n$ times the cost of
a function evaluation, a potentially significant expense.

When derivatives are not available from the simulation or through AD, an
alternative to ND is to
employ a \emph{derivative-free} optimization method \cite{Conn2009a}, that is,
one that relies only on evaluations of the function $F_i$ (or the aggregate
objective function, $f$). Because they are provided less information about
the objective, such methods
generally require a greater number of function evaluations than do
derivative-based methods.

\begin{figure}[tbh]
\begin{center}
\includegraphics[width=0.45\linewidth]{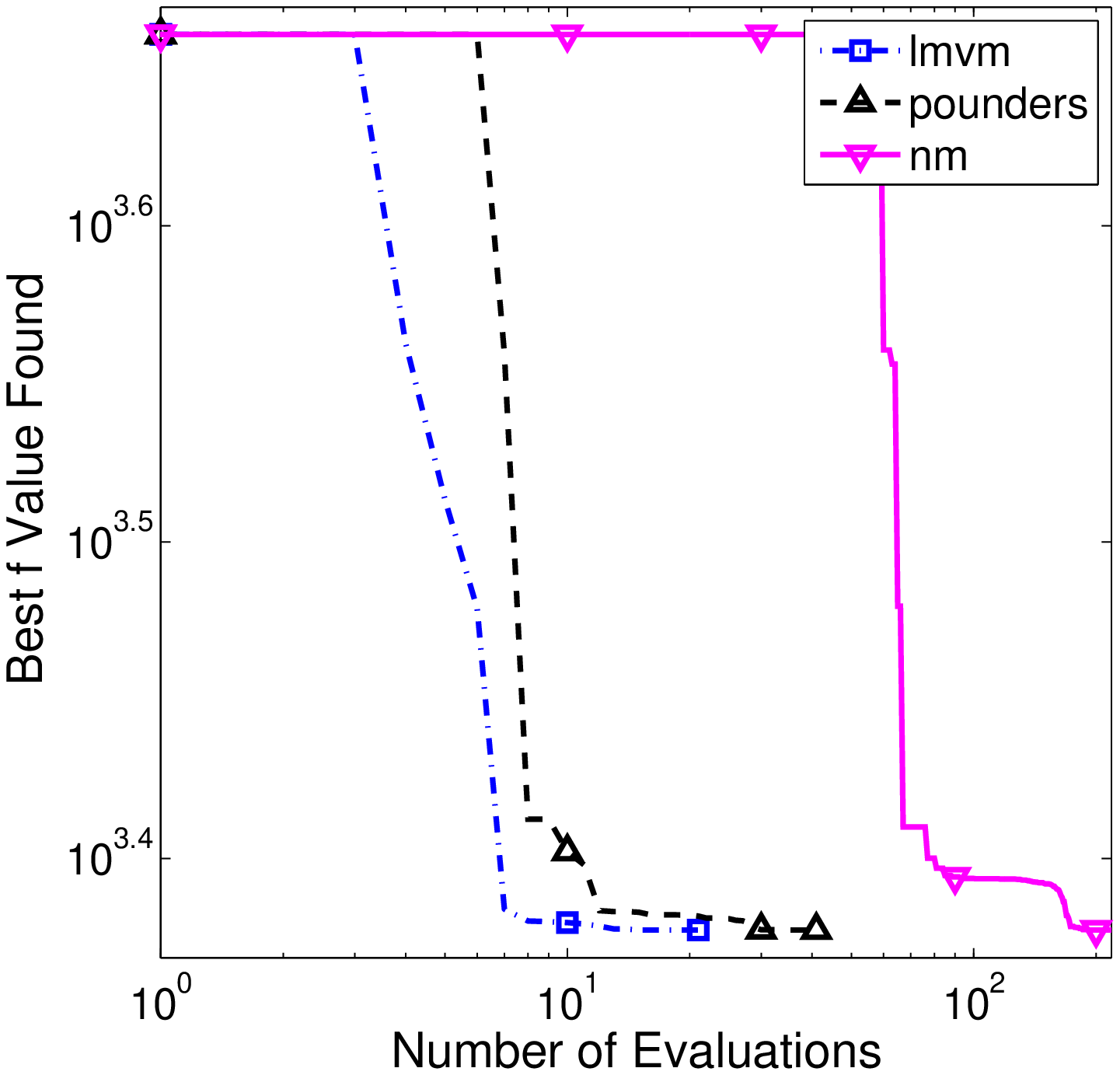} \hfill
\includegraphics[width=0.45\linewidth]{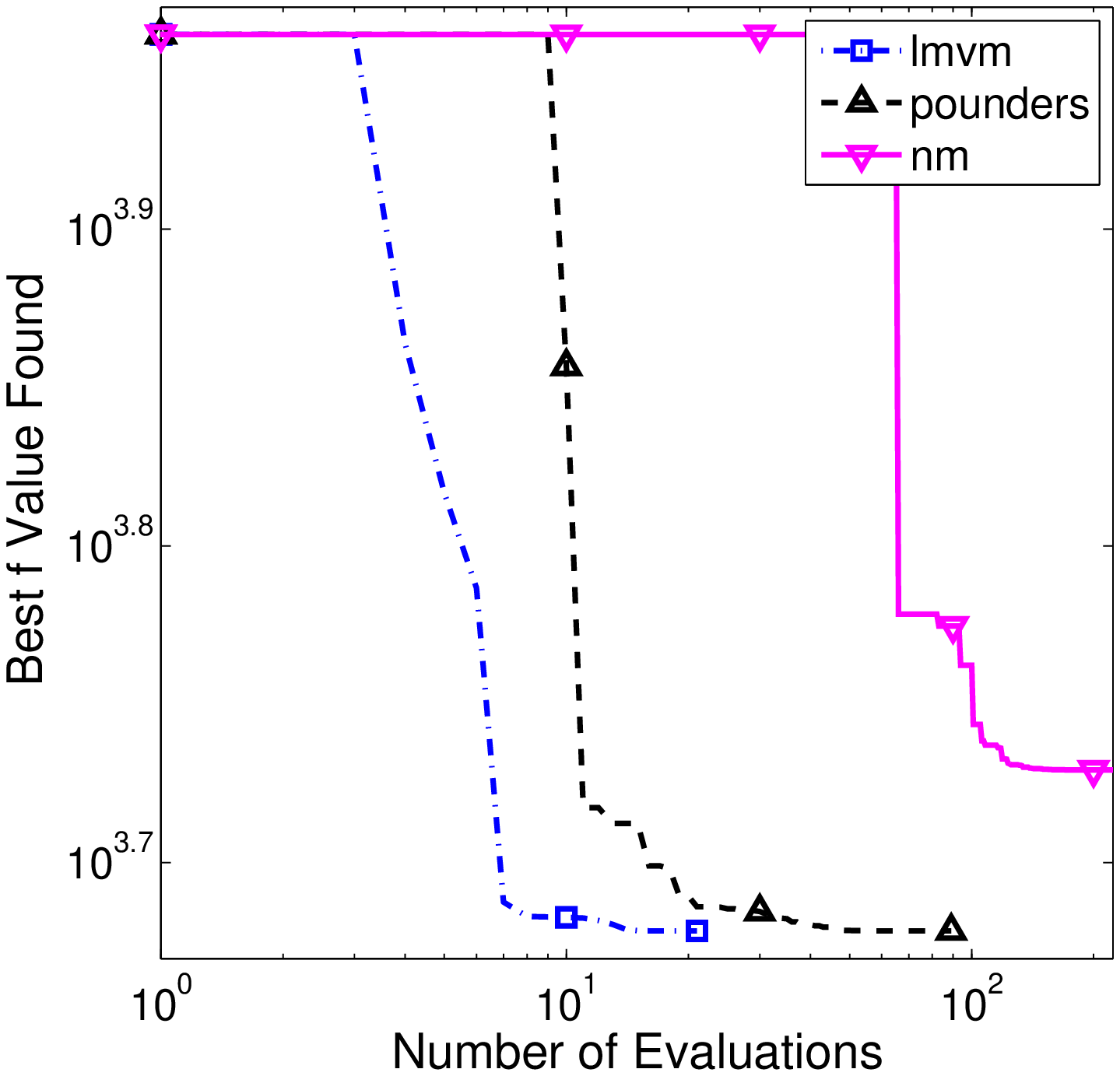}
\caption{Comparing the performance ($\log\log$ scale) of three \tao\ solvers
(limited-memory
variable metric, \pounders, Nelder Mead) on NLS test problems of the form
(\ref{eq:echwirut1}). The left plot is the original version
with $n_x=3$ variables; the right plot is the extended version with $n_x=6$
variables.
}
\label{fig:compare}
\end{center}
\end{figure}

We illustrate this concept by examining a typical example of an
unconstrained NLS problem, \texttt{chwirut1.c}. This example is included in the
Toolkit for Advanced Optimization (\tao), now a part of the Portable Extensible
Toolkit for Scientific Computation (\codes{PETSc}); see Section~\ref{sec:usage}.
This problem is based on the chwirut1 dataset \cite{chwirut1}, with the
extended version consisting of $n_x=6$ parameters and $n_d=428$ observables:
\begin{equation}\label{eq:chwirut1}
 \min_{\xb\in\R^6} f(\xb) = \sum_{i=1}^{214} \left(
\frac{e^{-x_1\nu_i}}{x_2 + x_3\nu_i} - d_i \right)^2 + \sum_{i=1}^{214} \left(
\frac{e^{-x_4\nu_i}}{x_5 + x_6\nu_i} - d_i \right)^2,
\label{eq:echwirut1}
\end{equation}
where $\{\nu_i: i=1,\dots,214\}$ are hyperparameters (metal distances) and
$\{d_i: i=1,\dots,214\}$ are experimental ultrasonic responses. The original
version does not include the second sum and thus has $n_x=3$ parameters and
$n_d=214$ observables.

We solved both versions using several of the algorithms available in \tao;
the results are shown in Figure~\ref{fig:compare}. The limited-memory
variable metric (LMVM) algorithm is a quasi-Newton method that utilizes first-derivative
information, and Nelder-Mead is a simplex-based derivative-free
method; neither method takes into account the sum-of-squares structure
present in (\ref{eq:nls}). \pounders\ is a derivative-free method that exploits
the availability of the residual vector $\Fb$ rather than just the single
aggregate $f$; we refer the reader to \cite{SWCHAP14} for a
mathematical description of the
algorithm. Figure~\ref{fig:compare} shows that when measured in terms of the
number of function evaluations, the derivative-based method LMVM reduces the
$f$ value considerably faster than do the derivative-free methods. If the
combined expense of a function and gradient evaluation is roughly the same as two
function evaluations, the advantage of LMVM over \pounders\ persists. However,
if the combined cost is roughly the same as $n_x+1$ function evaluations (as
would happen if using LMVM with gradients approximated by ND and forward
differences), then \pounders\ is faster.
In all these scenarios, the derivative-free method that does not exploit
the structure inherent in (\ref{eq:nls}) performs significantly worse.

\subsection{\pounders\ for Calibrating Energy Density Functionals}
Under the Universal Nuclear Energy Density Functional (UNEDF \cite{UNEDF13}) and Nuclear Computational Low-Energy Initiative (NUCLEI \cite{NUCLEI}) collaborations, a wide
variety of parameter estimation problems arose where derivatives of the
residuals were unavailable \cite{UNEDF0,UNEDF1,UNEDF2,BPW11,Ekstrom13}. Here we
focus on some of the results obtained when calibrating Skyrme energy density
functionals where HFBTHO \cite{HFBTHOv2} was the underlying simulator. As a
reminder, HFBTHO solves the Hartree-Fock-Bogoliubov equations for generalized
Skyrme functionals using the transformed harmonic oscillator basis under the
assumption of axial and time-reversal symmetry.
These built-in symmetries make HFBTHO particularly adapted to large-scale
surveys of nuclear properties and optimization problems
\cite{erler2012,stoitsov2009}.

\begin{table}[tb]
\caption{\label{tab:unedf0} Results for unconstrained and
bound-constrained derivative-free methods starting from SLy4 on the UNEDF0
problem ($n_x=12$, $n_d=108$, $n_N= 72$) \cite{UNEDF0}.}
\begin{indented}\item
\begin{tabular}{r||l|l|l|r}

 &  & \textbf{UNEDFnb} \cite{UNEDF0} & \multicolumn{2}{|c}{\textbf{UNEDF0} \cite{UNEDF0}} \\
\textbf{Method} & \codes{Nelder-Mead} & \pounders\  & \pounders\ (bounds)  &
$\sigb$ \\
\hline \hline

$\rho_c$ &   0.16155537 &   0.15104627  &  0.16052598  & 0.001 \\ 
$E^{NM}/A$ &   -16.115363 &   -16.063211  &   -16.05589 & 0.055 \\ 
$K^{NM}$ &    234.64613 & 337.87808  & \underline{230} & \textendash \\
$a^{NM}_{\mathrm{sym}}$ &    31.919478 & 32.454973 & 30.54294 & 3.058 \\ 
$L^{NM}_{\mathrm{sym}}$ &    46.186671 & 70.218532 & 45.080413 & 40.037\\ 
$1/M_s^*$ & 1.4306113 & 0.95727984 &  \underline{0.9} & \textendash \\ 
$C^{\rho \Delta \rho}_0$ & -78.133526 & -49.513502 & -55.260592 & 1.697 \\ 
$C^{\rho \Delta \rho}_1$ & 4.4779896 & 33.52886 & -55.622579 & 56.965 \\ 
$V^n_0$ & -240.42409 & -176.79601 & -170.37424 & 2.105 \\ 
$V^p_0$ & -252.81184 & -203.25488 & -199.20225 & 3.351 \\ 
$C^{\rho \nabla J}_0$ & -92.272157 & -78.456352 & -79.530829 & 3.423 \\ 
$C^{\rho \nabla J}_1$ & -27.615105 & 63.993115  & 45.63019 & 29.460 \\ 
\hline
$f(\xh)$ &    106.23493 &    41.865965 &      67.309821 & \\ 
$n_f$ & 300 & 268 &   300 & \\ 
\end{tabular}
\end{indented}
\end{table}

Table~\ref{tab:unedf0} summarizes the solutions obtained during
UNEDF0 computational experiments \cite{UNEDF0}. Each of the three runs was
started from the SLy4 parameterization \cite{chabanat1997,chabanat1998} and,
because of the computational expense of evaluating $n_d=108$
observables across $n_N=72$ even-even nuclei, run for a maximum of 300
evaluations. The first two columns represent the solutions from the Nelder-Mead and \pounders\ 
codes in \tao\ when solving the unconstrained problem, whereas the ``\pounders\ 
(bounds)'' column
shows the \pounders\ results when bound constraints (see
Table~\ref{tab:bounds}) are enforced for the 6 parameters that correspond to
nuclear matter properties, for which relatively strict constraints exist.
These bounds were added after it was noticed that the nuclear incompressibility
parameter in the unconstrained optimization had a large value that was
incompatible with experimental data.

\begin{table}[tb!]
\caption{\label{tab:bounds} UNEDF bound constraints and scaling intervals.}
\begin{indented}\item
\begin{tabular}{r||cc|cc|cc}
 & \multicolumn{2}{|c|}{\textbf{UNEDF0} \cite{UNEDF0}} &
\multicolumn{2}{|c|}{\textbf{UNEDF1} \cite{UNEDF1}} &
\multicolumn{2}{c}{\textbf{Scaling Bounds}}\\
 & & & \multicolumn{2}{|c|}{\textbf{UNEDF2} \cite{UNEDF2}}  \\
 &   $\lb$        & $\ub$    &   $\lb$  & $\ub$  & $\svb^l$ & $\svb^u$ \\
\hline
$\rho_c$ &   0.15        & 0.17    &   0.15  & 0.17  &0.14 & 0.18 \\
$E^{NM}/A$  &  -16.2       & -15.8   &  -16.2  & -15.8 & -17 & -15 \\
$K^{NM}$   &  \textit{190} & \textit{230} & \textit{220} & \textit{260} & 170 &
270 \\
$a^{NM}_{\mathrm{sym}}$ &   28        & 36   &   28  & 36 & 27 & 37  \\
$L^{NM}_{\mathrm{sym}}$ &  40        & 100  &  40   & 100 & 30 & 70\\
$1/M_s^*$ & 0.9         & 1.5    & 0.9     & 1.5  & 0.8 & 2.0 \\
$C^{\rho \Delta \rho}_0$ & $-\infty$ & $\infty$ &  $-\infty$ & $\infty$ & -100 & -40 \\
$C^{\rho \Delta \rho}_1$ & $-\infty$ & $\infty$ &  $-\infty$ & $\infty$ & -100 & 100 \\
$V^n_0$ &  $-\infty$ & $\infty$ &  $-\infty$ & $\infty$ &  -350 & -150 \\
$V^p_0$ &  $-\infty$ & $\infty$ &  $-\infty$ & $\infty$ &  -350 & -150 \\
$C^{\rho \nabla J}_0$ &  $-\infty$ & $\infty$ &  $-\infty$ & $\infty$ & -120 & -50 \\
$C^{\rho \nabla J}_1$ &  $-\infty$ & $\infty$ &  $-\infty$ & $\infty$ & -100 & 50 \\
\end{tabular}
\end{indented}
\end{table}

As seen from the number of function evaluations, $n_f$, performed, only
the unconstrained \pounders\ terminated short of the budget (because of a measure
of criticality, similar to $\|\nabla f(\xh)\|\leq \epsilon$, being satisfied);
however, the bound-constrained \pounders\ was also seeing negligible decreases
at the time the budget was exhausted. Since the bound-constrained problem
involves a smaller parameter space, the associated global minimum will
necessarily have a larger function value; this is borne out in the best
functions
values, $f(\xh)$, obtained by \pounders\ on these two problems. In contrast, as
with the test function in Figure~\ref{fig:compare}, the Nelder-Mead performance
is markedly worse. At the time of the UNEDF0 runs, each evaluation of $f$
required 12 minutes of wall time on 72 cores; thus each $300$-evaluation run
required 2.5 days.

For the bound-constrained problem, two of the $n_x=12$ parameters ($K^{NM}$ and
$1/M_s^*$) were active and hence restricted by the enforced bounds. Parameter
values that are active are \underline{underlined} in each of the tables in this
paper. We note that for subsequent studies, the bound on $K^{NM}$ was relaxed
based on this analysis; see Table~\ref{tab:bounds}. The final column in
Table~\ref{tab:unedf0} shows the
standard deviations $\sigb$ computed for each optimal parameter value; see
\cite{UNEDF0} for details of the computation of $\sigb$.

For UNEDF1, the number of nuclei and number of observables were increased, with
the resulting solution shown in the last column of Table~\ref{tab:unedf1}. A
similar run (the UNEDF1ex column) was performed with an additional parameter
$0 \leq \alpha_{ex} \leq 1$ multiplying the exchange Coulomb part of the
functional. This parameter was added with the intent of simulating many-body
correlation effects for the Coulomb term, and early work suggested it could
significantly improve reproduction of masses \cite{goriely2008}. The parameter
$\alpha_{ex}$ was treated as a free parameter (with bound
constraints corresponding to $[0,1]$). Adding
an additional parameter without increasing the amount of data should result in
an objective value no worse than when that parameter is held fixed. Although
this result cannot be guaranteed in practice when doing local optimization from
arbitrary starting points, Table~\ref{tab:unedf1} shows that this was indeed
the case for \pounders\ runs starting from UNEDF0. However, the improvement of
the fit was deemed too marginal to justify introducing an empirical parameter.
We note that moving from
UNEDF0 to UNEDF1, the active parameters changed; see the discussion in
\cite{UNEDF1}.

\begin{table}[htb]
\caption{\label{tab:unedf1} UNEDF1 ($n_d=115$, $n_N=79$) results obtained by
\pounders\ starting from the UNEDF0 parameterization. $\alpha_{ex}$ was fixed at
its nominal value of 1.0 for UNEDF0 and UNEDF1 and treated as a free parameter
(restricted to $[0, \, 1]$) in
UNEDF1ex.}
\begin{indented}\item
\begin{tabular}{r||l|lr}
 & \textbf{UNEDF1ex} \cite{UNEDF1} &  \textbf{UNEDF1} \cite{UNEDF1} & $\sigb$ \\
 \hline
$\rho_c$ &   0.15836673  &  0.15870677 & 0.00042 \\ 
$E^{NM}/A$ & \underline{-15.8} & \underline{-15.8} & \textendash \\ 
$K^{NM}$ &   \underline{220} &  \underline{220} & \textendash \\ 
$a^{NM}_{\mathrm{sym}}$ &    28.383952 &  28.986789 & 0.604 \\ 
$L^{NM}_{\mathrm{sym}}$ &  \underline{40} & 40.00479 & 13.136 \\ 
$1/M_s^*$ &    1.0018717  &  0.99242333 & 0.123 \\ 
$C^{\rho \Delta \rho}_0$ &   -44.601636  &  -45.135131 & 5.361 \\ 
$C^{\rho \Delta \rho}_1$ &   -180.95647  &  -145.38217 & 52.169 \\ 
$V^n_0$ &   -187.46859 & -186.0654 & 18.516 \\ 
$V^p_0$ &   -207.20942 & -206.57959 & 13.049 \\ 
$C^{\rho \nabla J}_0$ &   -74.339131  &   -74.026333 & 5.048 \\ 
$C^{\rho \nabla J}_1$ &   -38.837179  &  -35.658261 & 23.147\\ 
$\alpha_{ex}$ &    0.8135508 &  1.0 & \textendash \\ 
\hline
$f(\xh)$ &     49.341359 &     51.058424 \\ 
$n_x$ & 13  & 12 & \\ 
$n_f$ & 253  & 218 & \\ 
\end{tabular}
\end{indented}
\end{table}

\section{Consistency of Local Solutions}
\label{sec:starting}

As discussed in Section~\ref{sec:methods}, using local optimization methods has
the benefit of substantially reducing the number of expensive
simulations performed, when compared with global optimization methods. This
benefit, however, must be weighed against risks associated with being
dependent on the initial point from which a local run is started. We now
revisit some of the runs in the previous section and test the robustness of
\pounders\ under changes to the starting point, the simulation code, and the
data.

In each case, we find that \pounders\ obtains consistent (relative to
the original reported uncertainties) solutions.
Possible explanations of this (beyond being sufficiently
``lucky'') include the following
\begin{itemize}
 \item \pounders\ is relatively robust and tends to avoid getting stuck in poor
local minimizers.
\item The starting points are in reasonable parts of the
parameter space and
are thus conducive to yielding the same local minimizer/basin of attraction for
\pounders.
\item The data $\db$ and model $m$ result in an objective function
that is not very multimodal in this part of the parameter space.
\end{itemize}
We hypothesize that the likely reason is some combination of the above, but
these results  provide some confidence in the use of \pounders\ for this class
of problems.

\subsection{UNEDF0, revisited}

\begin{table}[tb]
\footnotesize
\caption{\label{tab:UNEDF0_v201} Rerun of \pounders\ on the UNEDF0
problem ($n_d=108$, $n_N=72$) using HFBTHO code (Ver 201) from two
different starting points.
The scaled difference columns represent the difference between the final value
found and the original UNEDF0 parameterization, scaled by the
uncertainties $\sigma_i$ in Table~\ref{tab:unedf0}.}
\begin{center}
\begin{tabular}{r||llr|llr}
& \multicolumn{2}{c}{\textbf{Starting from SLy4}} & \textbf{Scaled Diff.}
& \multicolumn{2}{|c}{\textbf{Starting from SKM*}} & \textbf{Scaled Diff.} \\ 
& \textit{initial} & \textit{final} & & \textit{initial} & \textit{final} \\ 
\hline
$\rho_c$ &     0.159539 & \same{0.160}486 &   -0.03954 &     0.160319 &
   \same{0.160}435 &   -0.09106 \\ 
$E^{NM}/A$ &     -15.9721 &     \same{-16.0}685 &     -0.2285 &          -16 &
   \same{-16.0}73 &    -0.3119 \\ 
$K^{NM}$ &      229.901 &          \underline{\same{230}} &  \textendash &      216.658 &          \underline{\same{230}} &            \textendash \\ 
$a^{NM}_{\mathrm{sym}}$ &      32.0043 &      \same{3}1.3393 &     0.2604 &
 30.0324 &      \same{3}1.7221 &     0.3856 \\ 
$L^{NM}_{\mathrm{sym}}$ &      45.9618 &      54.2493 &     0.2290 &
45.7704 &      60.4725 &     0.3844 \\ 
$1/,M_s^*$ &      1.43955 &          \underline{\same{0.9}} &   \textendash  &      1.26826 &          \underline{\same{0.9}} &            \textendash \\ 
$C^{\rho \Delta \rho}_0$ &     -76.9962 &     \same{-55.2}344 &    0.01545 &
-68.2031 &     \same{-55}.7348 &    -0.2794 \\ 
$C^{\rho \Delta \rho}_1$ &      15.6571 &     -64.1619 &    -0.1499 &
17.1094 &     -70.4274 &    -0.2599 \\ 
$V^n_0$ &      -285.84 &     \same{-170}.796 &    -0.2003 &         -280 &
\same{-170}.788 &    -0.1966 \\ 
$V^p_0$ &      -285.84 &     \same{-19}7.782 &     0.4238 &         -280 &
\same{-19}8.038 &     0.3474 \\ 
$C^{\rho \nabla J}_0$ &       -92.25 &     \same{-7}7.9436 &      0.4637 &
 -97.5 &     \same{-79}.2915 &    0.06990 \\ 
$C^{\rho \nabla J}_1$ &       -30.75 &      27.4519 &    -0.6171 &        -32.5
&      \same{4}9.5737 &      0.1339 \\ 
\hline
$f(\xh)$ & 1188.75 &      \same{67}.9034 & & 24814.1 &      \same{67}.5738 \\ 
$n_f$ & & 235 & & & 150 \\ 
\end{tabular}
\end{center}
\end{table}
\normalsize

The HFBTHO code has undergone several changes since the version used for the UNEDF0 optimization in
\cite{UNEDF0}. In particular, different initialization schemes of the HFB
problem have been implemented, the numerical accuracy of the direct Coulomb
potential has been improved, and a small bug on the rearrangement term for the
pairing field has been fixed; see
\cite{HFBTHOv2}. These changes result in minimal differences
 (often at the level of only a few keV on binding energies) to most observables
 used in the UNEDF0 calibration problem. Overall,
 the function value obtained at UNEDF0 is
roughly 67.985 for the latest version (Ver 201) of HFBTHO  compared with
67.310 (see Table~\ref{tab:unedf0}) for the version used in \cite{UNEDF0}.

Although small, these differences imply that the UNEDF0 parametrization no
longer satisfies the optimality conditions when computed with the new version of
HFBTHO. It is, therefore, natural to ask whether
additional optimization using this code version results in substantial changes.
In fact, this points to the general problem of the sensitivity of optimization
results on starting points: if one begins additional optimization starting from UNEDF0, or
SLy4, or any other starting point, will the resulting parameterization substantially differ from UNEDF0?

In Table~\ref{tab:UNEDF0_v201}, we report the results of the optimization obtained
from two very different starting points, the SLy4 parametrization of
\cite{chabanat1998}
used in our original UNEDF0 paper, and a starting point strongly inspired by the
SkM* parametrization of \cite{SKM*}. Since the binding energy per nucleon of SkM*
is out of our bounds, we fixed it arbitrarily at -16 MeV; similarly, SkM* does
not come with any prescription for pairing strengths, which we fixed at -280 MeV for both protons and
neutrons.
Table~\ref{tab:UNEDF0_v201} shows that in both cases---and despite SLy4 and
SkM*
being very different from UNEDF0 and from one another---similar solutions are
found. In fact, as the scaled difference column
($(\hat{x}^{\rm final}_i-\hat{x}^{\rm UNEDF0}_i)/\sigma_i$) shows, the two
solutions are both well within a single standard deviation of UNEDF0 (based on
the uncertainties $\sigb$ reported in Table~\ref{tab:unedf0}).

\subsection{UNEDF1, revisited}
We now consider the effect of changing the data $\db$ employed in the NLS
optimization of UNEDF1. We begin by motivating an estimate of the effect of
this change on the optimal parameter values $\xh$.

Formally, let $\xh\in \R^{n_x}$ minimize $f^{0}(\xb)=\|\Fb(\xb)\|_2^2$ as in
(\ref{eq:nls}). Now suppose that the residual  $\Fb(\xb)\in \R^{n_d}$ undergoes
a change by $\eb\in \R^{n_d}$, for example, because each normalized datum
$\frac{d_i}{w_i}$ is changed to $\frac{d_i}{w_i}+\epsilon_i$.
A second-order Taylor expansion of $f(\xb)=\|\Fb(\xb)+\eb\|_2^2$ about $\xh$ is
\begin{eqnarray*}
f(\xb) & \approx &
f(\xh) + 2 \eb^T\hat{J}(\xb-\xh) \\
& + & \frac{1}{2}(\xb-\xh)^T
\left(\nabla^2 f^0(\xh)
+2\sum_{i=1}^{n_d} \epsilon_i \nabla^2 F_i(\xh)\right) (\xb-\xh),
\label{eq:perturb}
\end{eqnarray*}
where $\hat{J}$ denotes the Jacobian matrix $[\frac{\partial
F_i(\xh)}{\partial x_j}]_{i,j}$ and we have used the first-order optimality
condition $\nabla f^0(\xh) =2 \hat{J}^T\Fb(\xh) = 0$.
When $\eb$ is small, this quadratic will be convex and hence minimized at
\begin{eqnarray*}
 \xb_{\epsilon} -\xh  &=& 2 \left(\nabla^2 f^0(\xh)
+ 2\sum_{i=1}^{n_d} \epsilon_i \nabla^2 F_i(\xh)\right)^{-1}  \hat{J}^T\eb \\
&=& 2\left(\nabla^2 f^0(\xh) \right)^{-1} \hat{J}^T\eb +\mathcal{O}(\|\eb\|^2).
\end{eqnarray*}
When $\Fb(\xh)$ is small, the Hessian $\nabla^2 f^0(\xh)$ is well-approximated
by $2\hat{J}^T\hat{J}$, which yields the approximation
\begin{equation}
\tilde{\xb}_{\epsilon} = \xh + \left(\hat{J}^T\hat{J} \right)^{-1}
\hat{J}^T\eb
 \label{eq:perturb1}
\end{equation}
of the new optimal solution for $f=\|\Fb(\xb)+\eb\|_2^2$.

We apply this estimate to the UNEDF1 problem when additional nuclear mass
data is added for the 17 new neutron-rich, even-even nuclei
measured in \cite{van_schelt_first_2013}. We refer to this new data as
the Argonne masses (AM); further details of the new observables can be found
in \cite[Supplementary material]{NUCLEIAM2014}.

\begin{figure}[tb]
\begin{center}
\includegraphics[width=0.5\linewidth]{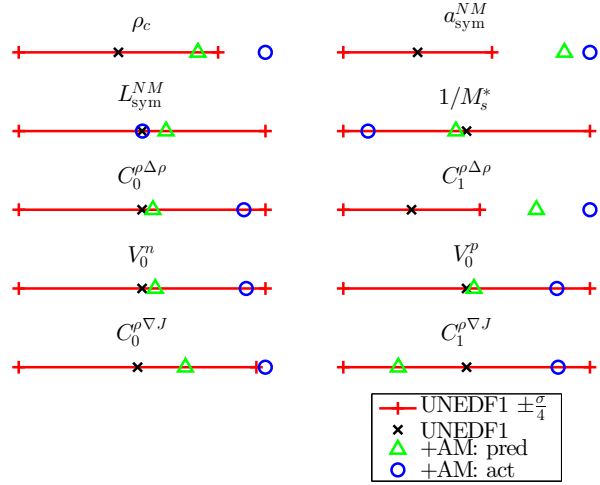}
\caption{(Half standard deviation) Intervals for the inactive parameters in
UNEDF1 \cite{UNEDF1}, the optimal parameters when the AM data is included as
predicted by (\ref{eq:perturb1}), and the actual optimal parameters found by
\pounders\ run from UNEDF1.}
\label{fig:AMSA}
\end{center}
\end{figure}

With the data vector $\db$ now containing $n_d=132$ components, we estimate the
effect of including the new observables by considering the vector $\eb\in
\R^{n_d}$ consisting of zeros, except in the 17 components corresponding to the
new observables. For these new observables, we take $\epsilon_i = F_i(\xh)$,
where $\xh$ is the UNEDF1 parameterization. For each of the 10 inactive
parameters of UNEDF1, Figure~\ref{fig:AMSA} illustrates the interval
corresponding to the UNEDF1 parameter value and a half standard deviation ($\pm
\frac{\sigma_i}{4}$, where $\sigb$ is reported in Table~\ref{tab:unedf1}).
The figure shows that the estimator (\ref{eq:perturb1}) predicts the new optimal
values to differ from UNEDF1 in only minor ways, each new value being within
$\frac{\sigma_i}{2}$ of UNEDF1. Also shown are the actual optimal values as
found by \pounders\ when the new data is included in an optimization begun from
UNEDF1 (see Table~\ref{tab:AM_data}). These actual values are also within
$\frac{\sigma_i}{2}$ of UNEDF1, with the predictions in (\ref{eq:perturb1})
generally indicating the correct direction of the change (with the exception
of $C_1^{\rho \nabla J}$).

Table~\ref{tab:AM_data} summarizes the solutions found by \pounders\ from two
different starting points.
Again, the solutions found are remarkably close to one another, each parameter
being with 0.07 of a standard deviation based (on the UNEDF1
uncertainties $\sigb$ reported in Table~\ref{tab:unedf1}).
Furthermore, as predicted by the parameters remaining close relative to their
uncertainties (Figure~\ref{fig:AMSA}), the $\chi^2$ values based on UNEDF1
($\frac{f(\xh)}{n_d-n_x}=\frac{51.058}{103}= 0.496$; see Table~\ref{tab:unedf1})
and the parameterization obtained with the AM data
($\frac{f(\xh)}{n_d-n_x}=\frac{54.01}{120} = 0.450$; see
Table~\ref{tab:AM_data}) are similar.

\begin{table}[tb]
\footnotesize
\caption{\label{tab:AM_data}
Reruns of the UNEDF1 optimization with the inclusion of 17 new AM data ($n_x=
12$, $n_d= 132$, $n_N=96$). The scaled difference columns are scaled by the
uncertainties $\sigma_i$ in Table~\ref{tab:unedf1}, the first column being the
difference between AM runs 1 and 2, the second column being the difference
between AM run 2 and the run without bound constraints enforced.}

\begin{center}
\begin{tabular}{r||ll|r||l|r}
 & \textbf{AM Run 1} & \textbf{AM Run 2} &  \textbf{Scaled
Diff.} & \textbf{No Bounds} &  \textbf{Scaled
Diff.} \\
\hline
start & UNEDF0 & UNEDF1 & & AM Run 2 & \\ 
 \hline
$\rho_c$ &   \same{0.1588}9255 &  \same{0.1588}6155 &
0.07381 & 0.15748674
& 3.273 \\ 
$E^{NM}/A$ &        \underline{\same{-15.8}} &   \underline{\same{-15.8}} &
- & -15.692799 & -\\ 
$K^{NM}$ &          \underline{\same{220}} &    \same{220.0}2317 &
- & 221.06558 & - \\ 
$a^{NM}_{\mathrm{sym}}$ &    \same{29.3}44856 &    \same{29.3}36203 &
0.01433 & 26.173927
&  5.236 \\ 
$L^{NM}_{\mathrm{sym}}$ &    \same{40.}714438 &    \same{40.}014867 &
0.05326 & 13.510725
& 2.018 \\ 
$1/M_s^*$ &   \same{0.96}859386 &    \same{0.96}78555 &
0.00600 & 0.91930059
& 0.3948 \\ 
$C^{\rho \Delta \rho}_0$ &   \same{-4}3.980091 &   \same{-4}4.028902 &
0.00910 & -39.479616
& -0.8395 \\ 
$C^{\rho \Delta \rho}_1$ &   \same{-11}4.29145 &   \same{-11}1.31777 &
-0.05700 & -150.49163
& 0.7509 \\ 
$V^n_0$ &   \same{-182.}23717 &   \same{-182.}15551 &
-0.00441 & -174.88812
& -0.3925 \\ 
$V^p_0$ &   \same{-20}3.98073 &   \same{-20}4.19083 &
0.01610 & -199.51881
& -0.3580 \\ 
$C^{\rho \nabla J}_0$ &   \same{-72.}417226 &   \same{-72.}668136 &
0.04970 & -71.753276
& -0.1812 \\ 
$C^{\rho \nabla J}_1$ &   \same{-3}2.920571 &   \same{-3}1.360678 &
-0.06739 & -31.708413
& 0.0150 \\ 
\hline
$f(\xh)$ &    54.0468 &    54.0140 &   & 46.3344 & \\ 
$n_f$ & 76 & 152 & & 74 & \\ 
\end{tabular}
\end{center}
\end{table}
\normalsize

The final columns in Table~\ref{tab:AM_data} return to the topic of
removing the bounds on the nuclear matter property parameters (recall the
UNEDF0 case in Table~\ref{tab:unedf0}). Here we see that $\chi^2$ can be
further reduced ($\frac{f(\xh)}{n_d-n_x}= 0.386$) if $E^{NM}/A$ and $K^{NM}$
are allowed to violate the bounds, but that the changes to the parameters are
substantial (up to 5 standard deviations for $a^{NM}_{\mathrm{sym}}$ alone),
even when starting from the bound-constrained solution ``AM Run 2.'' In
future work we plan to examine the effect of these bounds and the inclusion of
observables that better constrain the nuclear matter properties.

\section{Derivatives, Revisited}
\label{sec:ders}

The tables in Sections~\ref{sec:methods}~and~\ref{sec:starting} compare
differences in the initial and final values obtained after an optimization.
Although the computational budget used is indicated through the reported number
of function evaluations ($n_f$), these tables do not provide a sense of the
rate of progress made the reported algorithms. Were the majority of
the evaluations devoted to certifying approximate optimality? Or,
were substantial reductions of the objective obtained right up
until the final evaluations?

\begin{figure}[tb!]
\begin{center}
\includegraphics[width=0.5\linewidth]{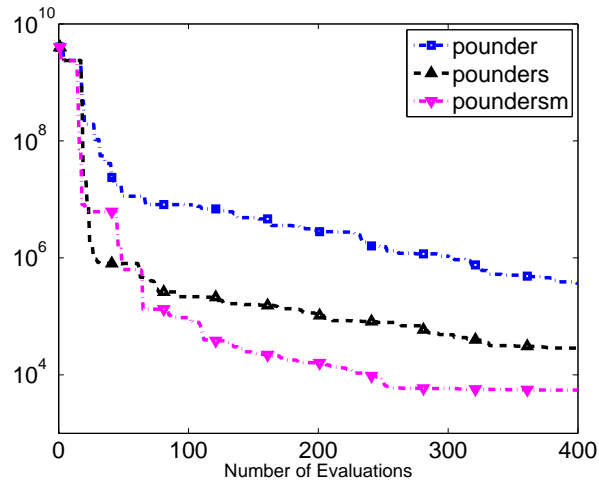}
\caption{Best $f$ value found as a function of the number of $f$ evaluations
for different model-based algorithms on a 17-dimensional parameter
estimation problem from \cite{BPW11}.}
\label{fig:someders}
\end{center}
\end{figure}

Figure~\ref{fig:someders} illustrates the rate of convergence on a
$n_x=17$-parameter problem involving the calibration of an occupation
number-based energy functional from \cite{BPW11}. The different methods in this
figure illustrate the benefits---in terms of convergence speed---of exploiting
structural knowledge about the optimization objective. All three methods are
based on the same model-based trust-region framework of \pounders; see
\cite{SWCHAP14}. The \codes{POUNDER} variant assumes that the optimization
algorithm does not have access to the residual vector and thus operates only
with $f$ values; \pounders\ uses the same formulation as in the previous
sections, whereby an entire residual vector $\Fb$ (in this case, consisting of
binding energies for $n_N=n_d=2049$ nuclei) is passed to the optimization
algorithm; and the \codes{POUNDERSM} variant exploits the fact that the
(first- and second-order) derivatives of each residual component are available
with respect to 3 of the 17 parameters.

Figure~\ref{fig:someders} can be placed into broader context by recalling
Figure~\ref{fig:compare}. As more residual derivatives are available, there is
a tendency to approach the derivative-based case (where the NLS structure is
exploited). In the other extreme, when only $f$ is available, the performance
of the \codes{POUNDER} variant is generally expected to be slightly better than
the Nelder-Mead code (see, e.g., \cite{JJMSMW09}).

\section{Using POUNDERS}
\label{sec:usage}

The \pounders\ algorithm can be a powerful tool for scientists
to evaluate and improve computationally expensive theoretical models so they
have better agreement
with experimental data.
We now outline some
of the typical requirements and usage of \pounders\ for solving applications
involving NLS problems.

The \pounders\ algorithm is included in the distribution of \codes{PETSc}/\tao,
an open-source software package developed at Argonne National Laboratory and
available
free of charge at \cite{petsc-web-page}. The software library can be built
on most common architectures and operating systems, using almost any modern
\texttt{C} compiler.

In order to use the \pounders\ algorithm, an objective function routine must be written
(in \texttt{C}/\texttt{C++} or \texttt{Fortran}/\texttt{Fortran90}) that can
separately compute each component, $F_i(\xb)$, of $\Fb(\xb)$ given a vector of
parameters $\xb$.
\pounders\ is a derivative-free method, so
no gradient information needs to be computed. In order to start the algorithm,
an initial set of parameters $\xb^0$ and an initial step length must also
be provided.

There are a number of features that can be used to improve
the performance and utility of \pounders.
One of these is the aforementioned enforcement of bound constraints. Finite
bounds can be provided for a subset (or all) of the parameters; these bounds
can be one-sided,  with only one of the lower or upper bound values being
finite. We note that \pounders\ assumes that these bound
constraints are \emph{unrelaxable}, meaning that the algorithm will never
attempt to evaluate the residual vector outside of the bounds. The benefit of
this restriction is that one can ensure that the underlying simulation is not
run in regions of parameter space where it may be error-prone or where its
output may not be defined. A limitation of this restriction is that, provided
the residual vector is well-behaved outside of these bounds, in some
cases a derivative-free algorithm requires fewer evaluations when these bounds
can be relaxed.

Whether finite bounds are provided or not, scaling of the variables is an
important consideration when calling \pounders. By default, \pounders\ 
fundamentally assumes that the objective $f$ experiences similar changes
under a unit change to each of the parameters. Consequently, we recommend that
the user apply an affine transformation to the parameters in the function that
\pounders\ calls.
For example, Table~\ref{tab:bounds} lists the scaling bounds used
throughout the HFBTHO-based optimizations reported in this paper. In the
layer between HFBTHO and \pounders, we apply a transformation
${\cal T}$ that maps the scaling rectangle in
Table~\ref{tab:bounds} to the unit hypercube, ${\cal T}([\svb^l,\svb^u])=
[0,1]^{n_x}$; the bound constraints being scaled by the same transformation.

Other features include the ability to initialize the internal model of the
application using precomputed parameter
sets and their objectives (warm-starting) to improve performance. There
are also
a number of features common to all {\codes{PETSc}/\tao} programs provided by the
\codes{PETSc}
framework; these include robust error handling, portability, command-line
argument parsing, and performance profiling \cite{petsc-user-ref}.

More detailed instructions for using \pounders\ are available from
\cite{petsc-web-page} or \cite{tao-man}, as well as example programs,
implementation details, and contact information.

\section{Summary}
\label{sec:summary}
In this paper we have examined optimization problem formulations that arise
when determining model parameters for nuclear energy density functionals. We
have stressed the potential multimodal nature of such problems and illustrated
the additional cost -- in terms of the number of model evaluations needed --
when the derivatives with respect to the parameters are unavailable.

Our numerical results in Sections~\ref{sec:methods}~and~\ref{sec:starting}
represent a significant empirical study of the solutions (parameter values
that approximately minimize the difference between theoretical models and
experimental data solutions)
found by the optimization solver \pounders\ in the recent Skyrme-based
functionals UNEDF0, UNEDF1, and UNEDF2.  We find that solutions obtained by
\pounders\ are remarkably robust to changes in the starting point used in the
optimization. Our results also show that the sensitivity analysis performed in
the development of these recent functionals is capable of predicting changes to
the optimal parameter values when new experimental data is included in the
optimization. We hope that the discussion of these problems and basic
description of the \pounders\ software will inspire the application of
numerical optimization methodologies in other areas of computational nuclear
physics.

\section*{Acknowledgments}
This work was supported by the SciDAC activity within
the U.S.\ Department of Energy, Office of Science, Advanced Scientific Computing
Research and Nuclear Physics programs under contract numbers DE-AC02-06CH11357
(Argonne) and DE-AC52-07NA27344 (Lawrence Livermore).
We gratefully acknowledge
high-performance computing resources operated by the
Laboratory Computing Resource Center at Argonne.

\section*{References}
\bibliographystyle{unsrt}
\bibliography{unedf}

\end{document}